\documentclass{raa}

\usepackage{graphicx,times}
\usepackage{natbib}
\usepackage{amssymb,amsmath}
\bibpunct{(}{)}{;}{a}{}{,}
\usepackage{ctex}
\AtBeginDocument{%
  \renewcommand{\abstractname}{Abstract}%
  \renewcommand{\figurename}{Figure}%
  \renewcommand{\tablename}{Table}%
  \renewcommand{\today}{\ifcase\month\or
    January\or February\or March\or April\or May\or June\or
    July\or August\or September\or October\or November\or December\fi
    \space\number\day, \number\year}%
}

\usepackage[pagebackref=true]{hyperref}

\begin{document}

\title{Magnetically Arrested Discs Powering Jets in a Large Sample of Low-Accretion FR I Radio Galaxies}

\volnopage{Vol.0 (202x) No.0, 000--000}
\setcounter{page}{1}

\author{Huibo~Fei
   \inst{1}
\and Han~He
    \inst{1}
\and Bei~You
   \inst{1,*}\footnotetext{$*$Corresponding author, e-mail: youbei@whu.edu.cn}
\and MinFeng~Gu
   \inst{2}
\and Liang~Chen
   \inst{2}
\and Xuheng~Ding
   \inst{1}
\and Leyi~Wang
   \inst{1}   
}

\institute{Department of Astronomy, School of Physics and Technology,
           Wuhan University, Wuhan 430072, China\\
      \and Shanghai Astronomical Observatory, Chinese Academy of Sciences, Shanghai 200030, China
\vs\no
{\small Received 202x month day; accepted 202x month day}}

\abstract{
We study a sample of 289 Fanaroff–Riley type I (FR I) radio galaxies selected from the LOFAR Two-Metre Sky Survey (LoTSS) DR1, identified by their edge-darkened radio morphologies. 
Using Sloan Digital Sky Survey (SDSS) DR17 optical photometry and spectroscopy, we derive Eddington-scaled accretion rates spanning -6.84 < log $\dot{m}$ < -0.87 (median $\approx$ -2.84). The vast majority of sources lie below $\dot{m} = 0.01$, indicating that their central engines are well described by advection-dominated accretion flows (ADAFs).
However, even for a rapidly spinning black hole with a = 0.95, the maximum jet power predicted by the Blandford–Znajek mechanism in the standard ADAF regime is lower than the observed jet power (estimated from 151 MHz radio luminosity) for approximately 70\% of the sample.
We demonstrate that the magnetically arrested disc (MAD) scenario — in which large-scale poloidal magnetic flux accumulates near the event horizon — can fully account for the powerful jets observed in these low-accretion systems. Within the MAD framework, the data are consistent with slow-spinning black holes with $a < 0.5$. This large, uniformly selected LoTSS sample extends the MAD requirement previously established for the bright 3CR FR I population, indicating that magnetically arrested discs are common in FR I radio galaxies across a wide range of luminosities.
}

\authorrunning{Fei et al.}
\titlerunning{MAD in FR\,I Radio Galaxies}

\maketitle

\section{Introduction}
\label{sect:intro}

Radio galaxies are radio-loud active galactic nuclei (AGN) in which relativistic jets are viewed at large angles to the line of sight, allowing their radio structures to be resolved from parsec to megaparsec scales. They are divided into Fanaroff–Riley type I and type II (FR I and FR II; \citealt{fanaroff1974}) according to the morphology of their radio lobes: FR Is exhibit edge-darkened emission, while FR IIs show edge-brightened structures.
Traditionally, FR II host galaxies were found to be more luminous than FR I hosts in the optical $R$ band \citep{ghisellini2001}. However, recent deep surveys have revealed a substantial overlap between the two classes in both radio and optical luminosity, except at the highest powers \citep{mingo2019}. The physical origin of this FR dichotomy remains one of the longest-standing puzzles in extragalactic astronomy and may involve differences in accretion mode, jet–environment interaction, or both (e.g. 
\citealt{gopal-Krishna1988,bicknell1995,baum1995,zirbel1995,ghisellini2001,cao2004,wu2011,best2012,li2014,kino2021,perucho2022}).

The connection between radio morphology and the central engine can be tested by comparing jet power with black hole (BH) mass. \citet{ghisellini2001} estimated BH masses from the absolute optical R-band magnitude of the host galaxy, while \citet{willott1999} calibrated the 151 MHz radio luminosity as a proxy for the time-averaged jet power (subject to substantial uncertainties associated with source age, environment, and the fraction of non-radiating particles). In this diagnostic framework, the ratio of jet power to BH mass provides a clear separation between FR I and FR II populations, suggesting that the morphological dichotomy may reflect fundamentally different accretion modes. Under this picture, FR I sources are expected to host advection-dominated accretion flows (ADAF;
\citealt{narayan1995}) with Eddington-scaled accretion rates (hereafter simply accretion rates) below $\sim$0.01, whereas FR II sources are typically powered by radiatively efficient standard discs (SSDs;
\citealt{shakura1973}). A more modern and physically motivated classification distinguishes low-excitation radio galaxies (LERGs) from high-excitation radio galaxies (HERGs; \citealt{best2012,heckman2014}). The vast majority of FR I sources belong to the LERG population, which is characterized by radiatively inefficient accretion.

This long-standing picture has been challenged by studies showing that both FR I and FR II morphologies can be produced by either SSDs or ADAFs
\citep{cao2004,hardcastle2018,hu2018,mingo2022}. The LOFAR Two-Metre Sky
Survey (LoTSS; \citealt{shimwell2017lofar}) provides a uniquely powerful dataset with which to revisit the FR dichotomy, owing to its combination of high sensitivity and excellent surface-brightness sensitivity. 
Using LoTSS Data Release 1 (DR1; \citealt{shimwell2019lofar}), \citet{mingo2019} uncovered a substantial population of low-luminosity, low-accretion-rate FR II galaxies and demonstrated a large overlap between FR I and FR II sources across the radio luminosity plane. They further proposed that host-galaxy stellar mass (or, more generally, environment) may play a decisive role in determining whether a low-power radio galaxy appears as an FR I or a low-luminosity FR II.

Powerful jets are also observed in some FR I radio galaxies that accrete at low rates. \citet{cao2004} examined optical and radio data for a sample of 33 FR I sources selected from the Third Cambridge Catalog (3C) by \citet{chiaberge1999}. They calculated the maximum jet power extractable via the Blandford–Znajek mechanism \citep{blandford1977} from a rapidly spinning black hole ($a = 0.95$) surrounded by an ADAF, adopting this spin value as a conservative upper limit. Surprisingly, more than one-third of the sources exceeded this theoretical ceiling. Crucially, these high-jet-power objects do not possess systematically higher black hole masses than the remainder of the sample. This discrepancy demonstrates that jet-production physics in FR I radio galaxies is more complex than the simplest ADAF picture \citep{ghisellini2001} and strongly motivates the search for additional mechanisms that can enhance jet power at low accretion rates.

Similar low-accretion systems with powerful jets are observed not only in FR I radio galaxies but also in other radio-loud AGN and in the well-studied cases of M87 and Sgr\,A$^{\ast}$ \citep{zamaninasab2014,eht2021,eht2022,yuan2022}. A natural explanation is that the magnetic flux threading the black hole has been underestimated in standard models. In the magnetically arrested disc (MAD) scenario \citep{narayan2003}, large-scale poloidal magnetic flux is advected inward and accumulates near the event horizon until magnetic pressure becomes dynamically important, partially arresting the inflow. Strong observational support for this process comes from the 2018 outburst of the black hole X-ray binary MAXI J1820+070, during which \citet{You2023Sci...381..961Y} detected radio and optical lags of 8 and 17 days relative to the X-ray emission; the $\sim$ 8-day radio delay was interpreted as the timescale for MAD formation. Beyond X-ray binaries, \citet{li2024hergmad} found that the jet efficiencies and multiwavelength properties of 3CRR quasars and high-excitation radio galaxies are broadly consistent with MAD expectations, suggesting that magnetic-flux saturation may also operate in radiatively efficient radio-loud AGN.General-relativistic magnetohydrodynamic simulations further demonstrate that MADs can launch jets with efficiencies $\eta_{\rm jet}\approx {\rm few}\times 100$ per cent
\citep{tchekhovskoy2011}, far exceeding the available accretion power and thereby providing a viable mechanism for powerful jets in radiatively inefficient flows \citep{zdziarski2015,lisl2022}.

\citet{he2024magnetically} applied this MAD framework to a sample of 17 FR I radio galaxies from the 3CR catalog and demonstrated that it successfully accounts for their unexpectedly powerful jets. This finding raises the important question of whether magnetically arrested discs are a common feature of FR I systems more generally. To address this question, we now extend the analysis to a much larger and more representative sample of 289 FR I galaxies selected from the LOFAR Two-Metre Sky Survey (LoTSS) Data Release 1 \citep{mingo2019}.

In this paper, we present a comprehensive study of the accretion properties and jet powers of this LoTSS FR I sample. Section 2 describes the sample selection, black hole mass estimation, and multiwavelength observations. In Section 3, we compare the observed jet powers with theoretical predictions from the Blandford–Znajek mechanism under both standard ADAF and MAD conditions. Section 4 provides independent verification of the low accretion rates using multiple methods and discusses the implications for black hole spin.
The cosmological parameters $H_{0} = 70~\rm km~s^{-1}~Mpc^{-1}$ and
$\Omega_{\rm m} = 0.3$ are adopted throughout this paper.

\section{Sample and Data}
\label{sect:data}

\citet{mingo2019} classified the LoTSS DR1 radio-galaxy catalog using the LoMorph algorithm followed by visual inspection. From this catalog, we selected all sufficiently extended and well-resolved FR I sources, yielding an initial parent sample of 1281 objects. We then cross-matched these sources with the Sloan Digital Sky Survey Data Release 17 (SDSS DR17; \citealt{abdurro2022seventeenth}) and imposed the following requirements: (i) a reliable spectroscopic redshift, and (ii) successful two-dimensional structural decomposition of the SDSS g-band image using GALIGHT \citep{2020ApJ...888...37D,2021JOSS....6.3283B}that cleanly separates the unresolved nuclear component from the host galaxy. These selection criteria reduced the sample to 289 FR I galaxies. The main reasons for exclusion were the lack of an SDSS counterpart ($\sim$ 65.18\% of the parent sample) and failure of the GALIGHT decomposition for morphologically complex or dust-obscured hosts ($\sim$ 12.25\%).

The BH masses are estimated from the SDSS DR17 \citep{abdurro2022seventeenth}
optical magnitudes, using the host-galaxy magnitude--BH mass relation of
\citet{mclure2002},
\begin{equation}
    \log(M_{\rm BH}/{\rm M_{\odot}}) = -0.50(\pm0.02)M_{\rm R}-2.96(\pm0.48)
\end{equation}

To obtain AGN luminosities in the $g$ band, we perform two-dimensional
structural decomposition of the SDSS images with GALIGHT. The host-galaxy light is modeled with multiple components, such as a bulge, disc, and
unresolved point source, allowing the nuclear component to be separated from
the host emission \citep{2020ApJ...888...37D,2021JOSS....6.3283B}.
\begin{figure}
    \centering
    \includegraphics[width=1\textwidth]{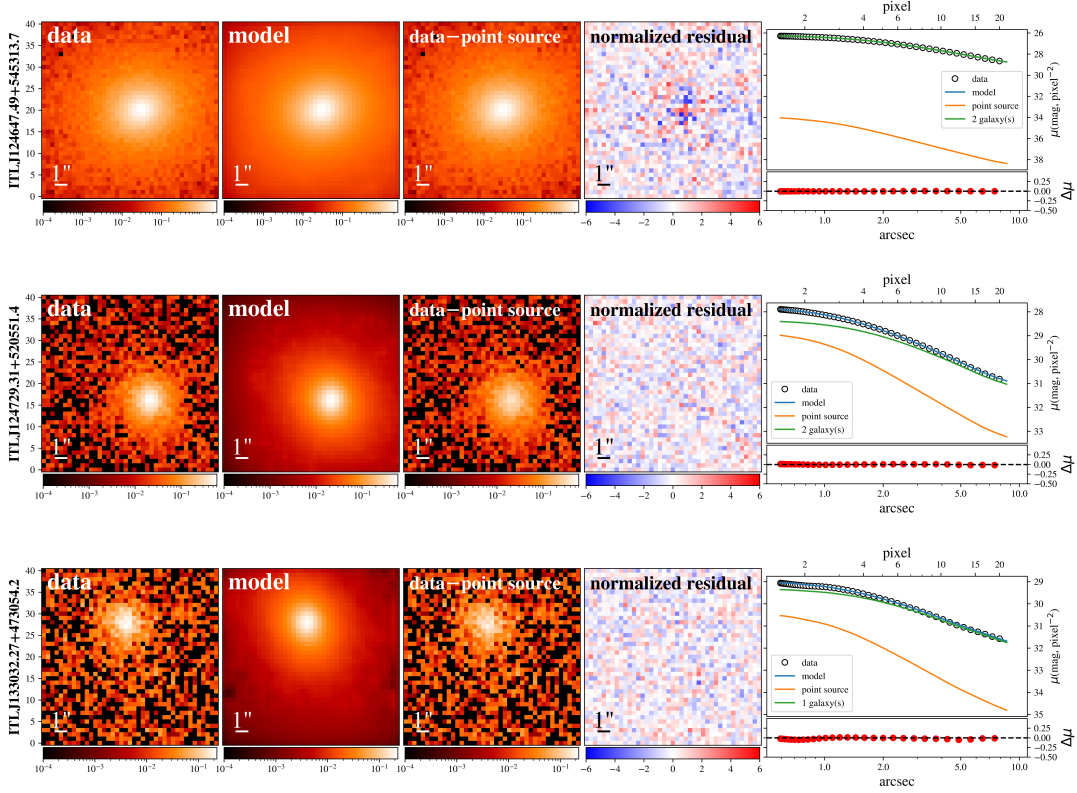}
    \caption{Two-dimensional decomposition of SDSS g-band images for three representative FR I galaxies with different optical luminosities using GALIGHT. From left to right: (1) observed image; (2) best-fit model (AGN point source + host galaxy); (3) residual after subtracting the AGN component (i.e., pure host galaxy); (4) normalized residual (data minus model divided by variance); (5) one-dimensional surface brightness profiles (top) and residuals (bottom). Open circles denote the data, the blue line the total model, the orange line the AGN, and the green line the extended host. The 1-D profiles show the surface brightness including the data (open circles), the best-fit model (blue line), the AGN (orange line), and the model for the extended sources (green line, i.e., host galaxy and other objects). Note that the 1-D surface brightness profiles are shown for illustration purposes only; the actual fitting is performed on the two-dimensional images.}

\end{figure}

\begin{table}
 \begin{center}
  \caption{Representative entries from the final sample of 289 FR I radio galaxies. Columns are: (1) source name in LoTSS DR1; (2)–(3) right ascension and declination; (4) redshift from SDSS spectroscopy; (5) logarithm of jet power (with $f$ = 1); (6) logarithm of black hole mass; (7) rest-frame logarithm of optical core luminosity ($\alpha$ = 0.5 K-correction); (8) logarithm of Eddington-scaled accretion rate.}
\label{table_sample}
\begin{tabular}{lccccccc}
\hline\noalign{\smallskip}
Name & RA & DEC & $z$ & $\log Q_{\rm jet}$ & $\log M_{\rm BH}$ & $\log L_c$ & $\log\dot{m}$\\
     &    &     &     & (erg\,s$^{-1}$) & ($M_\odot$) & (erg\,s$^{-1}$) & \\
\hline\noalign{\smallskip}

ITLJ123530.80+522828.9 & 188.878 & 52.475 & 1.653 & 44.23 & 10.90 & 46.03 & $-$1.976 \\
ITLJ145603.67+504826.2 & 224.015 & 50.807 & 1.379 & 43.91 & 10.14 & 45.42 & $-$1.815 \\
ITLJ113251.06+541031.8 & 173.213 & 54.176 & 1.626 & 44.93 & 10.22 & 45.54 & $-$1.783 \\
ITLJ124834.18+512806.4 & 192.142 & 51.468 & 0.351 & 43.08 & 9.01  & 44.58 & $-$1.534 \\
ITLJ121548.85+522447.2 & 183.954 & 52.413 & 1.515 & 44.25 & 9.52  & 44.98 & $-$1.634 \\
ITLJ131407.16+503219.1 & 198.530 & 50.539 & 0.484 & 42.16 & 8.18  & 43.32 & $-$1.970 \\
ITLJ141419.55+491820.4 & 213.581 & 49.306 & 0.568 & 43.24 & 8.47  & 44.25 & $-$1.320 \\
ITLJ145515.90+480814.3 & 223.816 & 48.137 & 1.326 & 44.27 & 9.80  & 45.02 & $-$1.877 \\
ITLJ120644.89+464336.6 & 181.687 & 46.727 & 0.842 & 43.87 & 8.78  & 44.48 & $-$1.400 \\
ITLJ121509.96+462715.4 & 183.791 & 46.454 & 0.720 & 44.37 & 10.06 & 45.60 & $-$1.566 \\
ITLJ133437.19+563147.6 & 203.655 & 56.530 & 0.343 & 43.41 & 8.62  & 44.43 & $-$1.293 \\
ITLJ144611.40+484615.2 & 221.548 & 48.771 & 1.200 & 43.66 & 10.04 & 45.14 & $-$2.004 \\
ITLJ110615.75+552707.3 & 166.566 & 55.452 & 0.993 & 44.50 & 10.04 & 45.41 & $-$1.724 \\
ITLJ135056.26+511438.6 & 207.734 & 51.244 & 0.434 & 42.86 & 8.17  & 43.50 & $-$1.777 \\
ITLJ144057.33+510619.6 & 220.239 & 51.105 & 1.636 & 44.16 & 10.19 & 45.48 & $-$1.819 \\
ITLJ144953.31+535029.6 & 222.472 & 53.842 & 0.361 & 42.43 & 7.98  & 43.43 & $-$1.641 \\
ITLJ150835.92+514923.1 & 227.150 & 51.823 & 0.520 & 42.70 & 8.15  & 44.44 & $-$0.809 \\
ITLJ140248.77+491350.6 & 210.703 & 49.231 & 0.498 & 42.35 & 8.44  & 43.13 & $-$2.411 \\
ITLJ133203.62+525735.0 & 203.015 & 52.960 & 2.250 & 44.62 & 9.52  & 44.86 & $-$1.753 \\
ITLJ142741.62+520410.1 & 216.923 & 52.069 & 1.305 & 43.49 & 9.09  & 44.52 & $-$1.677 \\

\noalign{\smallskip}\hline
\end{tabular}
\end{center}
\end{table}

Jet power can be estimated in several ways, including radio-lobe emission under the equipartition assumption \citep{willott1999}, the radio core-shift effect \citep{shabala2012}, spectral modeling \citep{ghisellini2014}, the total energetic output divided by the source age \citep{mahatma2019}, and X-ray cavity measurements \citep{birzan2008}. Following \citet{he2024magnetically}, we estimate the jet power with the widely used scaling relation of \citet{willott1999},
\begin{equation}
    Q_{\rm jet} \simeq 3\times 10^{38} f^{3/2} L_{151}^{6/7}~\rm W
\end{equation}
where $L_{151}$ is the total radio luminosity at 151\,MHz in units of ${\rm 10^{28}\,W\,Hz^{-1}\,sr^{-1}}$. We use the radio fluxes measured by LoMorph from \citet{mingo2019}, adopt a power-law spectral index of $\alpha=0.7$ for the $K$-correction, and convert the observed radio luminosity to the 151 MHz band. {\it LoMorph} \citep{mingo2019} recovers low-surface-brightness extended emission through a flood-filling algorithm, providing a more complete characterization of the source. This makes the method particularly reliable for FR I galaxies. However, the primary aim of \citet{mingo2019} was morphological classification rather than precise flux calibration. We therefore adopt a conservative 50\% uncertainty for all 151 MHz measurements as an upper bound on the error budget. Propagating this uncertainty through Equation (2) yields a logarithmic uncertainty of $\sim 0.19$ dex in $Q_{\rm jet}$, which has negligible impact on our main results. The factor $f$ encapsulates uncertainties in source age, environment, magnetic field strength, and the ratio of radiating to non-radiating particles, and is usually taken to lie in the
range $1\leq f\leq20$. For FR I systems such as M87, \citet{cao2004} found
that $f=1$ gives a jet power consistent with previous estimates. We therefore
adopt $f=1$ as our fiducial value and revisit this choice in Section~\ref{sect:compare}.

\section{Results}
\label{sect:results}

\subsection{Accretion Rates of the LoTSS FR I Sample}
\label{sect:rate}

We estimate the dimensionless accretion rate $\dot{m}=\dot{M}/\dot{M}_{\rm Edd}$, in which
$\dot{M}_{\rm Edd}=L_{\rm Edd}/(\eta_{\rm eff}c^2)$ and $L_{\rm Edd}$ is the Eddington luminosity. Under this assumption, $\dot{m}$ is equivalent to the Eddington ratio,
$\lambda = L_{\rm bol}/L_{\rm Edd}$, where $L_{\rm bol}$ is the bolometric
luminosity.

To estimate $L_{\rm bol}$, we use the relation between absolute $B$-band
magnitude and bolometric luminosity derived by \citet{mclure2004} from 372
sources in the 2dF 10K quasar catalogue \citep{croom2001} and the SDSS Quasar
Catalogue II \citep{schneider2003}:
\begin{equation}
    \label{eq:Mb&bol}
    M_{\rm B} = -2.66(\pm 0.05)\log [L_{\rm bol}/\rm W]+79.36(\pm 1.98)
\end{equation}
This calibration can be expressed in the commonly used form
\citep{Kaspi2000ApJ...533..631K,wu2013}:
\begin{equation}
        \label{eq:Lbol}
    L_{\rm bol}~({\rm erg~s^{-1}}) = 10L_{\rm B},
\end{equation}
where $L_{\rm B}$ is the luminosity at 4400\,\AA.

\begin{figure}
    \centering
    \includegraphics[width=0.6\textwidth]{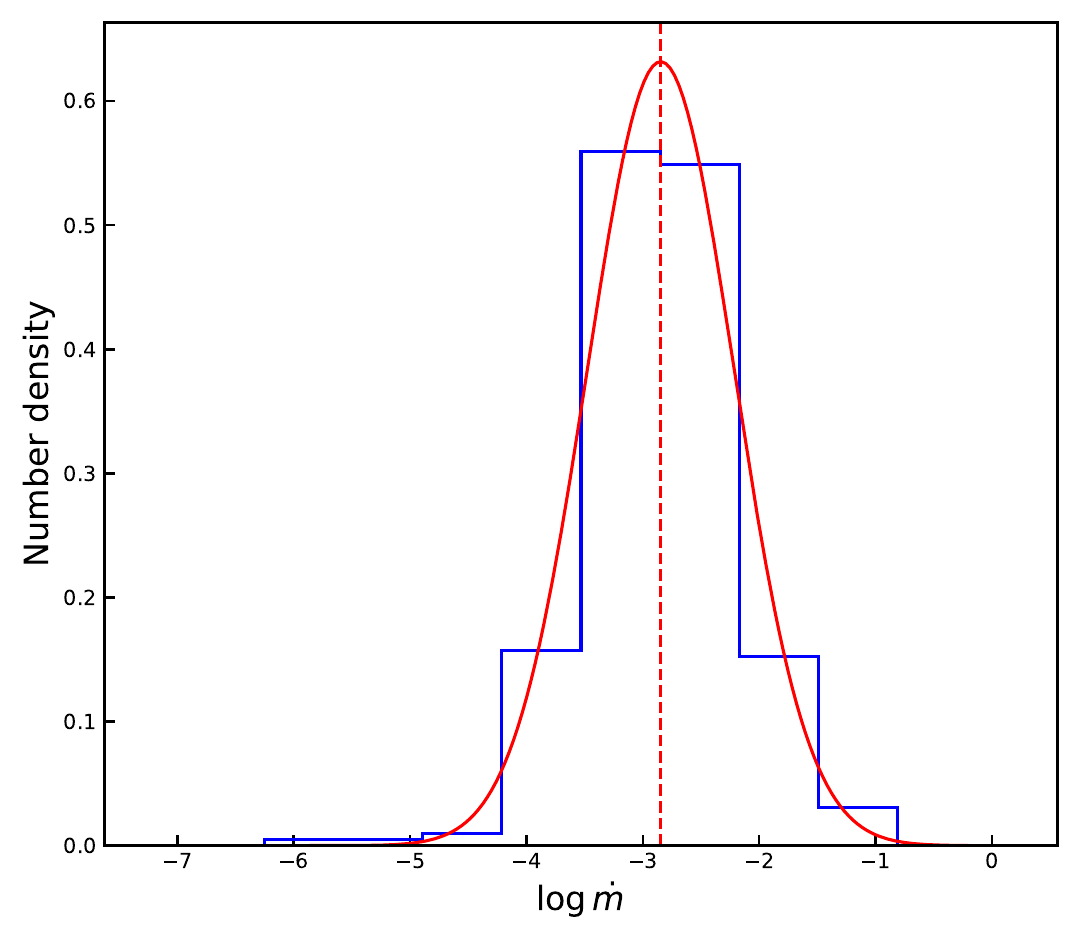}
    \caption{
    Distribution of Eddington-scaled accretion rates ($\log \dot{m}$) for the 289 FR I galaxies. Bin size is 0.5. The red histogram shows the observed distribution, the solid red curve is the best-fit Gaussian ($\mu=-2.84$ and $\sigma=0.64$), and the pink dashed line indicates the mean.}
    \label{fig:rate}
\end{figure}

We use the GALIGHT-derived $g$-band core luminosity $L_g$ as a proxy for $L_{\rm B}$; the central wavelength of the SDSS $g$ band is 4720\,\AA. The resulting bolometric luminosities span
$39.03<\log L_{\rm bol}<46.03$. Combining these luminosities with the BH masses described above, we obtain Eddington-scaled accretion rates in the range $-6.84<\log \dot{m}<-0.87$ (Fig.~\ref{fig:rate}). The vast majority of sources lie below $\dot{m}=0.01$, the conventional division between low- and high-accretion regimes, implying that the most objects are of ADAF-like accretion. We model the distribution of $\log\dot{m}$ by a Gaussian profile (pink line in Fig.~\ref{fig:rate}), finding a mean of $\mu=-2.84$ and a standard deviation of $\sigma=0.64$.

\subsection{Jet Power Deficit in the Standard ADAF Scenario}
\label{sect:ADAF}
Using these sample parameters, we calculate the maximum jet power that can be extracted from a spinning BH surrounded by an ADAF via the Blandford--Znajek mechanism \citep{livio1999}, adopting the mean accretion rate
$\log \dot{m} = -2.84\pm0.64$ as representative. The maximum power extracted from a spinning BH is
\begin{equation}
        L_{\rm BZ}=\left(\frac{B^2}{4\pi}\right) \pi R_{\rm h}^2
    \left(\frac{R_{\rm h} \Omega_{\rm h}}{c} \right)^2c,
\label{eq:L_bz}
\end{equation}
where $R_{\rm h}$ is the horizon radius and $\Omega_{\rm h}$ is the
angular velocity of the BH (both depend on spin $a$ and mass
$M_{\rm BH}$. Following \citet{cao2004}, we assume the magnetic pressure in an ADAF scales as $p\sim B^2/8\pi$. Consequently, $L_{\rm BZ}$ depends on $\dot{m}$, $a$, and $M_{\rm BH}$.

\begin{figure}
    \centering
    \includegraphics[width=0.6\textwidth]{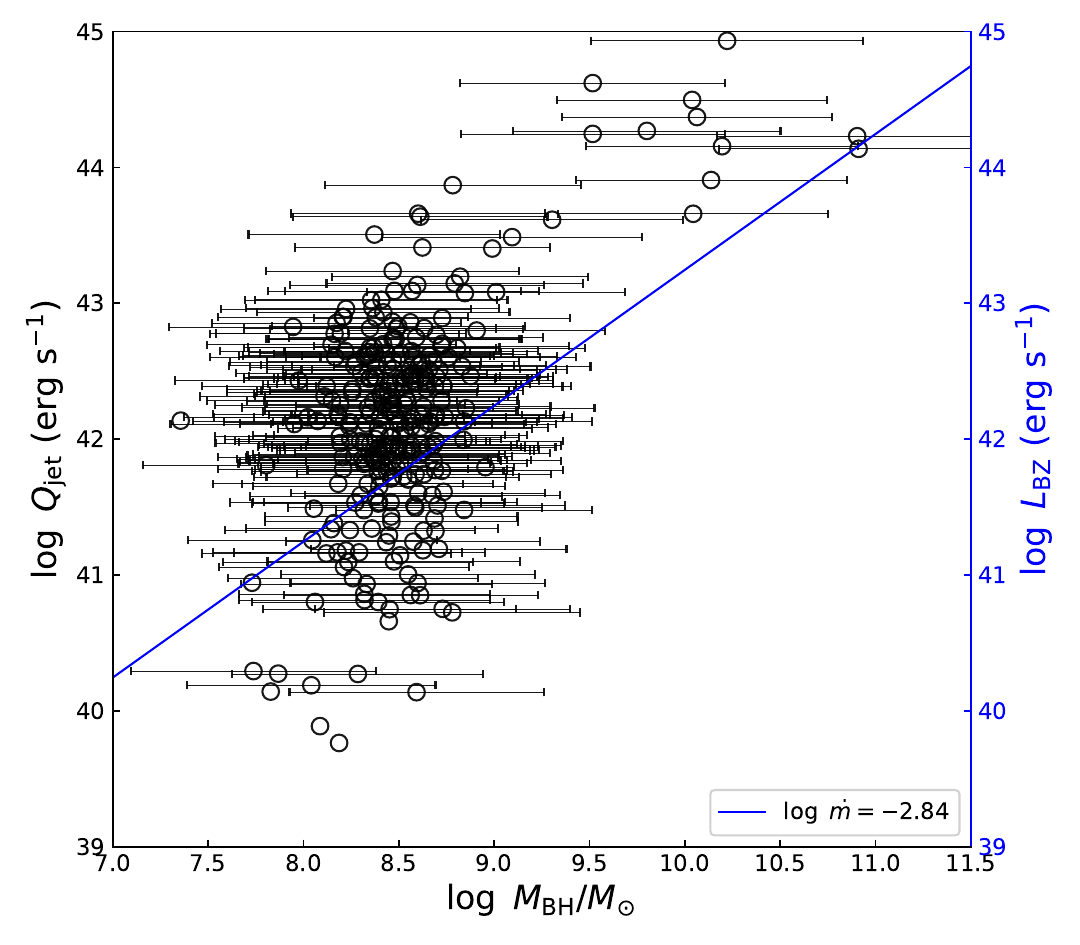}
    \caption{
    Jet power versus black hole mass for the full sample. Open circles show individual sources, with horizontal error bars indicating the uncertainty in black hole mass. The blue line and shaded region show the maximum jet power predicted by the Blandford--Znajek mechanism for an ADAF around a black hole with spin $a = 0.95$ at the median accretion rate $\log \dot{m} = -2.84 \pm 0.64$. The spin $a = 0.95$ is adopted as an upper limit. The uncertainty in $\log(M_{\rm BH}/M_\odot)$ is taken as approximately $\pm 0.5$ dex.}
    \label{fig:adaf_model}
\end{figure}

Figure~\ref{fig:adaf_model} compares the observed jet powers (open circles) with the maximum jet power predicted by the Blandford–Znajek mechanism for an ADAF around a rapidly spinning black hole ($a=0.95$; blue line and shaded region). Even with this choice of spin (which provides an upper envelope), approximately 70\% of the sources lie above the predicted line at the median accretion rate $\log \dot{m}=-2.84$. This demonstrates that the standard ADAF model significantly underestimates the jet power in the majority of our sample.

\subsection{Jet Power Enhancement in the MAD Scenario}
\label{sect:mad}

As shown in Figure~\ref{fig:adaf_model}, the majority of sources have $\dot{m}<10^{-2}$ and thus reside in a radiatively inefficient
accretion regime. Nevertheless, their observed jet powers systematically exceed the ADAF-based Blandford--Znajek estimate. In the BZ mechanism, the jet is powered by a spinning BH threaded by a large-scale magnetic field, so the jet power is sensitive to the magnetic-field strength at the horizon \citep{livio1999}. The large discrepancy, therefore, implies that the magnetic field in the inner accretion flow is stronger than in the standard ADAF estimate.

Powerful jets at low accretion rates are also observed in other radio-loud AGN as well as in M87 and Sgr\,A$^{\ast}$ \citep{zamaninasab2014,eht2021,eht2022,yuan2022,kuze2025}. To explain the required stronger magnetic field, we consider the magnetically arrested disc model \citep[MAD;][]{narayan2003,yuan2014,Dihingia2024}. In this scenario, large-scale poloidal magnetic flux is advected inward and accumulates near the event horizon \citep{ljw2025}. Recent simulations show that the dynamical structure of outflows driven by such large-scale magnetic fields is consistent with the MAD framework \citep{ljw2022}. 
Global GRMHD simulations of tilted MADs further demonstrate that electromagnetic torques generally act to align the disc with the black hole spin, although in retrograde configurations this alignment is countered by hydrodynamic torques, leading to persistent misalignment and solid-body precession \citep{jiang2025,gupta2026}.
As the magnetic field grows, its pressure can become strong enough to oppose the inflow and trap a large amount of magnetic flux near the BH, thereby partially arresting the inflow. We therefore re-estimate the jet power by allowing for magnetic-flux saturation at the horizon.

The accumulated magnetic field arrests the accretion flow at a magnetospheric radius $R_{\rm m}$. The region inside $R_{\rm m}$ constitutes a MAD, in which the strong magnetic field provides significant support against gravity \citep{narayan2003}. The field strength can be written as
\begin{equation}
        B_{\rm MAD}\sim 1.5\times10^9(1-f_\Omega)^{1/2}\epsilon^{-1/2}
    m_{\rm BH}^{-1/2}\dot{m}^{1/2}R^{-5/4}~\rm G.
\label{eq:b_mad}
\end{equation}
Here $\epsilon=v_{\rm R}/v_{\rm K}\sim 0.01$--$0.1$ is the ratio of radial to Keplerian velocity in the MAD, and $f_\Omega=\Omega/\Omega_{\rm K}$, where $\Omega$ and $\Omega_{\rm K}$ are the angular velocity of the accretion flow and the Keplerian angular velocity, respectively. In Equation~(\ref{eq:b_mad}), $m_{\rm BH}\equiv M_{\rm BH}/M_\odot$ is the dimensionless black hole mass, $\dot{m}$ is the dimensionless accretion rate, and $R$ is the dimensionless radius in units of the Schwarzschild radius $R_{\rm S}=2GM_{\rm BH}/c^2$. We adopt fiducial values $\epsilon=0.01$, $f_{\Omega}=0.5$, and $R\sim R_{\rm ISCO}$, where $R_{\rm ISCO}$ is the innermost stable circular orbit expressed in units of $R_{\rm S}$ and depends on the BH spin \citep{bardeen1972,you2012}:
\begin{equation}
R_{\rm ISCO}=\frac{1}{2}\left\{3 + z_2 - \left[ (3-z_1)(3+z_1+2z_2) \right]^{1/2}\right\}.
\label{eq:r_isco}
\end{equation} 
with $z_1=1 + (1-a^2)^{1/3} [ (1+a)^{1/3} + (1-a)^{1/3} ]$ and
$z_2=(3a^2+z_1^2)^{1/2}$. Substituting Equations~(\ref{eq:b_mad})
and~(\ref{eq:r_isco}) into Equation~(\ref{eq:L_bz}) gives the MAD jet power as follows:
\begin{equation}
        L_{\rm BZ}=5.625\times10^{17}(1-f_{\Omega})\epsilon^{-1}
    m_{\rm BH}^{-1}\dot{m}R_{\rm ISCO}^{-5/2}
    \frac{R_{\rm h}^4\Omega_{\rm h}^2}{c}.
\label{eq:final_Lbz}
\end{equation}

\begin{figure}
    \centering
    \includegraphics[width=0.6\textwidth]{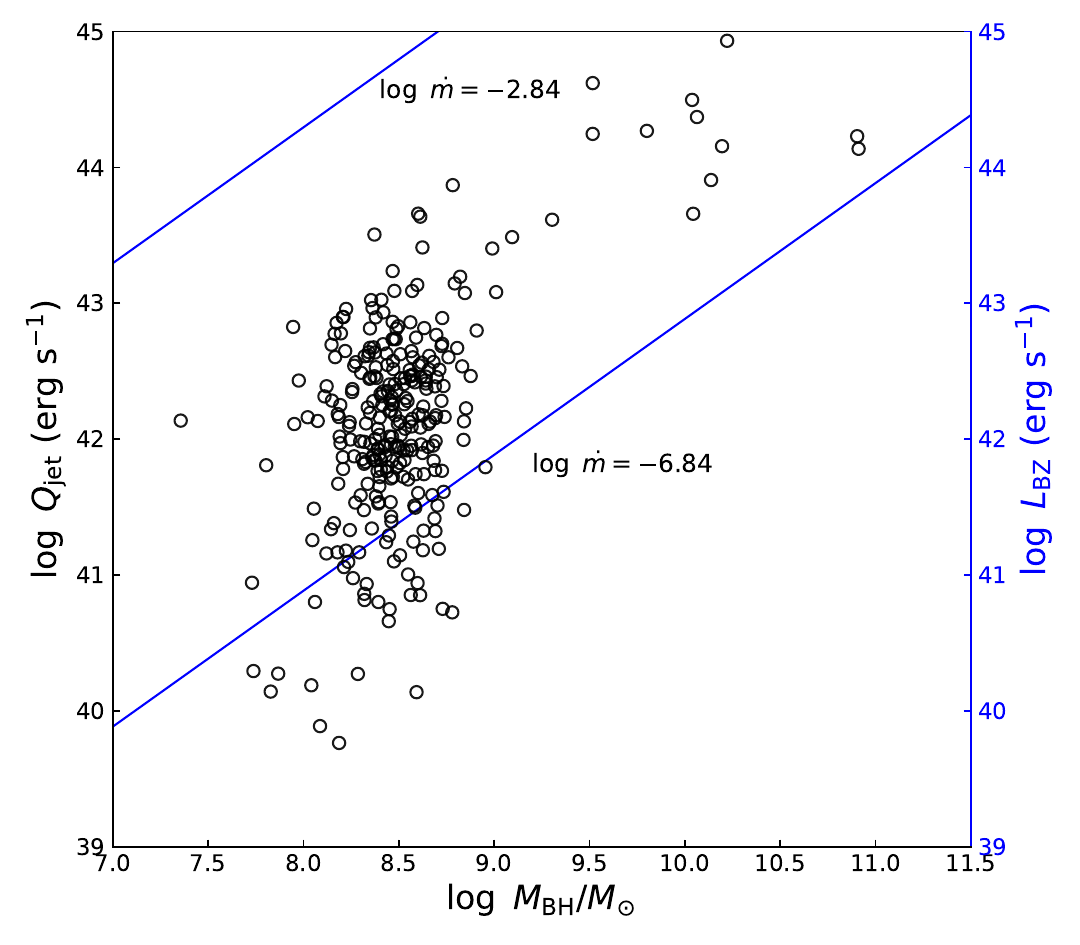}
    \caption{
    Jet power versus black hole mass. Blue solid lines show the MAD-predicted maximum jet power (Equation 8) for two representative accretion rates ($\log \dot{m}=-6.84$ and $-2.84$) at spin $a = 0.95$.The MAD model fully encompasses the observed data points. }
    \label{fig:mad_0.95}
\end{figure}

As demonstrated in Section~\ref{sect:rate}, our sample spans a wide range of low Eddington-scaled accretion rates. In Fig.~\ref{fig:mad_0.95}, we show the MAD-predicted jet power versus BH mass for two representative accretion rates ($\log \dot{m}=-6.84$ and $-2.84$) at $a=0.95$. The MAD tracks fully encompass the observed data points, demonstrating that magnetic flux accumulation near the event horizon can readily account for the powerful jets observed in our sample.

\section{Discussion}
\label{sect:discuss}

For the 289 FR I radio galaxies in our final sample, the estimated
Eddington-scaled accretion rates are mostly below $10^{-2}$, with
$-6.84<\log\dot{m}<-0.87$ (median $\approx -2.84$). These results indicate that the sample is dominated by radiatively inefficient, low-accretion sources. Given these low accretion rates, the observed jet powers cannot be explained by standard ADAFs and instead require strong magnetic-field accumulation near the event horizon, as predicted by the MAD scenario.

\subsection{Other Methods for Accretion Rate}
\label{sect:rate2}

In addition to the photometric measurements, we make use of derived
spectroscopic quantities from the MPA-JHU value-added catalogue \citep{brinchmann2004stellar}. Emission-line measurements in this catalogue are obtained after subtraction of the underlying stellar population using the method of \citet{tremonti2004origin}. We employ these measurements to obtain an independent estimate of the bolometric luminosity in Section~\ref{sect:rate2}. The catalogue also provides stellar masses \citep{kauffmann2003stellar}, the 4000\,\AA\ break strength (D4000), and the stellar velocity dispersion ($\sigma$), all of which are used in the present analysis.

We employ two additional methods to verify the accretion-rate estimates. First, we estimate the bolometric luminosity from the nuclear H$\beta$ luminosity using the empirical relation between H$\beta$ luminosity and the corresponding absolute $B$-band magnitude:
\begin{equation}
        \label{eq:Hb&Mb}
    \log L_{\rm H\beta} = (-0.34 \pm 0.012)M_{\rm B},
    + (35.1\pm0.25)
\end{equation}
where the H$\beta$ luminosities are taken from the MPA-JHU value-added
catalogue. Combining Equations~(\ref{eq:Hb&Mb}) and~(\ref{eq:Mb&bol}), we obtain bolometric luminosities in the range
$10^{40}$--$10^{46}$\,erg\,s$^{-1}$, with corresponding Eddington ratios from $10^{-6}$ to $10^{-2}$. Second, we estimate the BH mass independently from the stellar velocity dispersion using the $M_{\rm BH}$--$\sigma$ relation of \citet{mcconnel2013}:
\begin{equation}
    \log \left( \frac{M_{\rm BH}}{M_\odot} \right)
= 8.32 + 5.64 \log \left( \frac{\sigma}{200 \,\mathrm{km\,s^{-1}}}, \right)
\end{equation}
where $\sigma$ is the stellar velocity dispersion. As in
Section~\ref{sect:rate}, we compute the Eddington luminosity from the BH mass and the bolometric luminosity from the $g$-band core luminosity. The resulting Eddington ratios span $10^{-6}$--$10^{-3}$. Both independent estimates yield accretion rates that are comparable to, or lower than, the fiducial estimates in Section~\ref{sect:rate}. These results confirm that the sample is in a low-accretion state and strengthen the tension with the standard ADAF jet-power prediction.

In addition, we perform spectral decomposition for a subset of the sample with the \texttt{PyQSOFit} package \citep{Guo2018ascl}, following the methodology of \citet{Shen2019}. The optical spectra are taken from SDSS DR17 \citep{abdurro2022seventeenth}. Host-galaxy emission is modeled using the decomposition scheme implemented in \texttt{PyQSOFit} \citep{Ren2024}. The resulting accretion-rate estimates are consistent in order of magnitude with those derived in Section~\ref{sect:rate}.

\begin{table}
 \begin{center}
  \caption{
  SDSS spectroscopic identifiers and key measurements for a subset of the sample. Columns: (1) source name; (2)–(4) SDSS plate, MJD, and fiberID; (5) stellar velocity dispersion (km\,s$^{-1}$); (6) H$\beta$ flux ($10^{-17}$\,erg\,s$^{-1}$\,cm$^{-2}$). All values in columns (5) and (6) are taken from the MPA-JHU value-added catalog.}
\label{table1}
\begin{tabular}{lccccc}
\hline\noalign{\smallskip}
Name & plate & MJD & fiberID & $V_{\rm disp}$ & H$\beta$ flux \\
     &       &     &         & (km\,s$^{-1}$) & ($10^{-17}$\,erg\,s$^{-1}$\,cm$^{-2}$) \\
\hline\noalign{\smallskip}

J124834.18+512806.4 & 886  & 52381 & 221 & 284.98 & 50.94  \\
J133437.19+563147.6 & 1320 & 52759 & 31  & 400.00 & 259.71 \\
J135056.26+511438.6 & 1669 & 53433 & 622 & 95.46  & 9.82   \\
J123913.86+553628.9 & 1020 & 52721 & 623 & 256.18 & 2.09   \\
J110330.17+512656.3 & 877  & 52353 & 456 & 292.20 & 14.26  \\
J134545.42+533254.6 & 1042 & 52725 & 592 & 363.76 & 312.98 \\
J135735.87+483829.0 & 1670 & 53438 & 122 & 186.64 & 3.10   \\
J125326.39+505429.4 & 1279 & 52736 & 572 & 256.66 & 16.78  \\
J132435.15+504103.7 & 1667 & 53430 & 12  & 183.91 & 10.03  \\
J134655.85+505505.3 & 1669 & 53433 & 507 & 236.42 & 4.81   \\
J151245.03+523242.9 & 1165 & 52703 & 566 & 212.31 & 1.15   \\
J135521.45+460843.4 & 1466 & 53083 & 575 & 248.75 & 18.99  \\
J130906.90+500608.1 & 1281 & 52753 & 365 & 238.27 & 1.97   \\
J133238.66+513147.8 & 1668 & 53433 & 503 & 217.84 & 0.26   \\

\noalign{\smallskip}\hline
\end{tabular}
\end{center}
\end{table}

\begin{figure}
    \centering
    \includegraphics[width=0.85\textwidth]{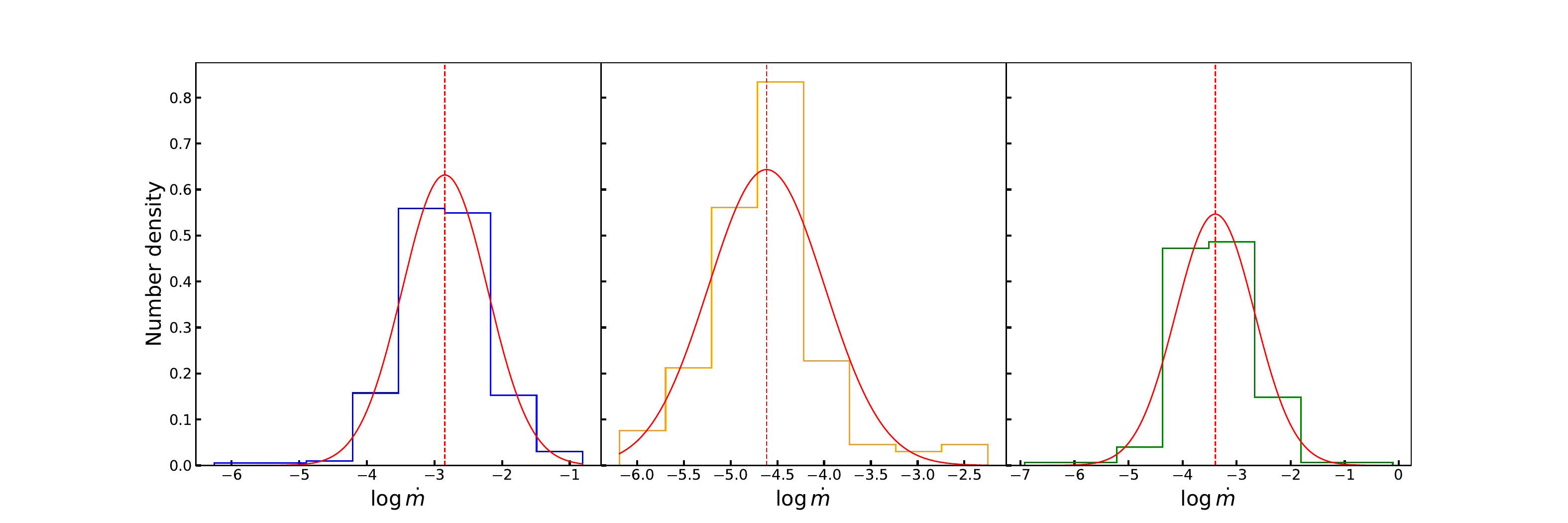}
    \caption{
    Comparison of accretion-rate distributions derived from three independent methods. Blue: g-band core luminosity (Section 3.1); red: H$\beta$-based bolometric luminosity; green: stellar velocity dispersion + g-band luminosity. Dashed lines mark the mean of each distribution. All three methods consistently show that the vast majority of sources have $\log \dot{m} < -2$.}
    \label{fig:rate2}
\end{figure}

\subsection{Estimates for Black Hole Spins}
\label{sect:spin}

\begin{figure}
    \centering
    \includegraphics[width=0.6\textwidth]{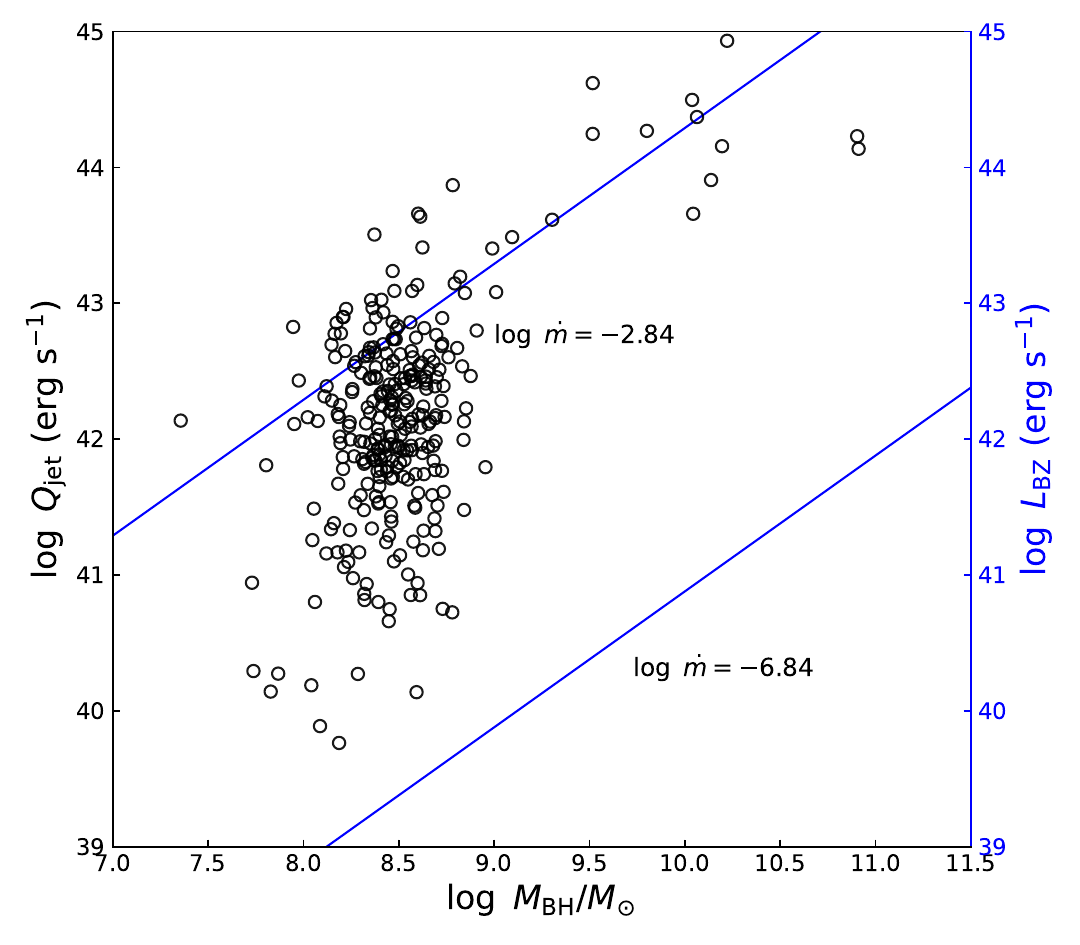}
    \caption{
    Jet power versus black hole mass for a slowly spinning black hole ($a$ = 0.1). Blue solid lines show the MAD-predicted maximum jet power (Eq. \ref{eq:final_Lbz}) at $\log \dot{m}=-6.84$ and $\log \dot{m}=-2.84$.Most sources are consistent with the $a=0.1$ MAD prediction, while a small fraction of high-power sources remain above the model tracks.}
    \label{fig:mad_0.1}
\end{figure}

In addition to accretion rate and BH mass, the jet power depends strongly on BH spin \citep{livio1999,narayan2003}. In Section~\ref{sect:mad}, we adopted a high spin of $a=0.95$ to estimate the maximum MAD jet power. Here we also consider a slowly spinning BH with $a=0.1$. As shown in Figure~\ref{fig:mad_0.1}, most sources still lie above the predicted MAD jet power for $a=0.1$, indicating that very low spins cannot account for the observed jets in the sample.

\begin{figure}
    \centering
    \includegraphics[width=0.6\textwidth]{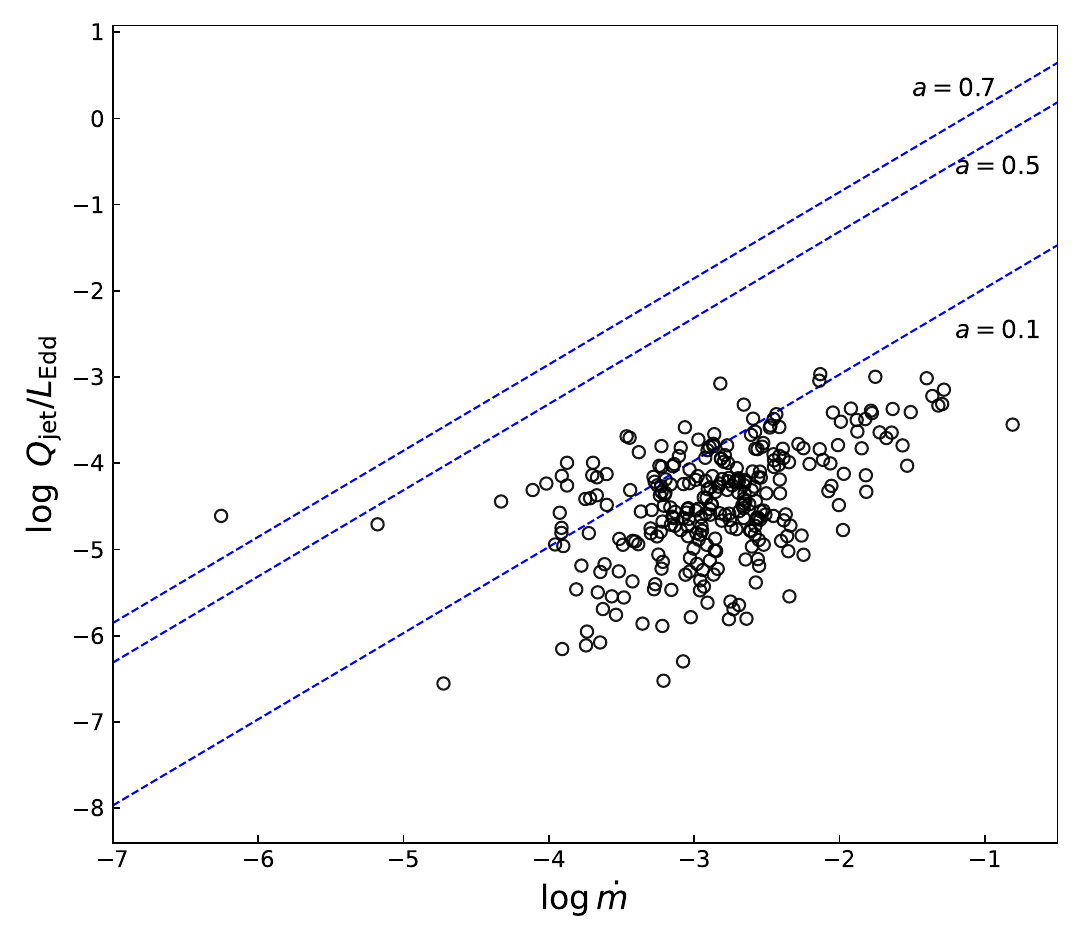}
    \caption{
    Jet efficiency ($Q_{\rm jet}/L_{\rm Edd}$) versus Eddington-scaled accretion rate. Dashed lines show the MAD-predicted maximum efficiency for black hole spins $a$ = 0.7, 0.5, and 0.1 (Eq. \ref{eq.spinsimulation}). Most sources lie between the $a$ = 0.5 and $a$ = 0.1 tracks, suggesting moderate spins.}
    \label{fig:rate_Q_Ledd}
\end{figure}

Moreover, the maximal estimated jet power scales with the BH mass after substituting
$R_{\rm h}=\dfrac{GM_{\rm BH}}{c^2}\left[1+\left(1-a^2 \right)^{1/2}\right]$
and $\Omega_{\rm h}=\dfrac{ca}{2R_{\rm h}}$ into
Equation~(\ref{eq:final_Lbz}), which gives:
\begin{equation}
        L_{\rm BZ}={\rm C}\,m_{\rm BH}\dot{m}R_{\rm ISCO}^{-5/2}a^2
    \left[1+\left(1-a^2 \right)^{1/2} \right]^2
    \label{eq.spinsimulation}
\end{equation}
where ${\rm C} \simeq 1.41\times10^{17}(1-f_{\Omega})\epsilon^{-1}G^2c^{-3}$.
We therefore show in Fig.~\ref{fig:rate_Q_Ledd} the relation between accretion rate and $Q_{\rm jet}/L_{\rm Edd}$ for $a=0.7$, 0.5,
and 0.1, respectively. Most sources fall under the predicted lines for  $a=0.1$, suggesting that the BHs in the sample are likely to be low spinning under our fiducial assumptions.

We also perform a Spearman rank correlation test between the inferred spin and the accretion rate, yielding a correlation coefficient of $r=-0.4$ with significance of $\sim 7\sigma$. This indicates a statistically significant but moderate anti-correlation. Therefore, the inferred spin values should be interpreted with caution, accounting for uncertainties in both the accretion rate and the jet power.

\subsection{Comparison of These Samples with Other Potential MAD Sources}
\label{sect:compare}

\begin{figure}
    \centering
    \includegraphics[width=0.6\textwidth]{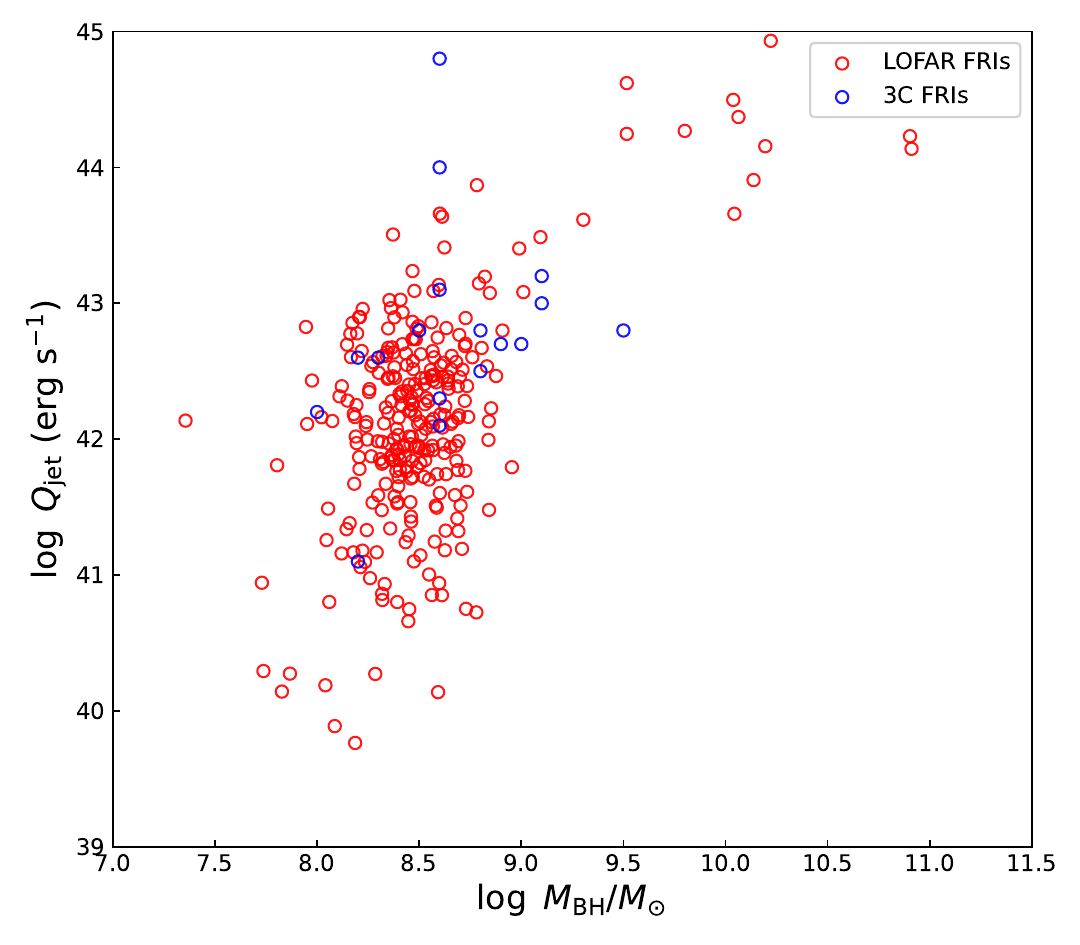}
    \caption{
    Comparison of jet power versus black hole mass between our LoTSS sample (red circles) and the 3CR FR I sample from \citet{he2024magnetically} (blue circles). The two samples lie in a broadly similar region of the $Q_{\rm jet}$-$M_{\rm BH}$ plane, indicating that magnetically arrested discs are a common feature across FR I populations at different luminosities.}
    \label{fig:compare}
\end{figure}

\citet{he2024magnetically} studied FR I radio galaxies in the original 3C catalogue and reached a similar conclusion: MADs provide a natural explanation for powerful jets in low-accretion FR I systems. Our LoTSS sample occupies a broadly similar region of the $Q_{\rm jet}$--$M_{\rm BH}$ plane (see Fig. \ref{fig:compare}), suggesting that the larger LoTSS-selected sample shares the same underlying jet physics as the classical 3C FR I population.

\citet{he2024magnetically} further argued that the BHs in the 3CR sample are moderately or rapidly spinning, with $a\gtrsim0.5$. Our fiducial analysis favors somewhat lower, moderate spins, but this difference should not be over-interpreted. The jet power depends sensitively on the uncertain factor
$f$ in $Q_{\rm jet} \simeq 3\times 10^{38} f^{3/2} L_{151}^{6/7}~\rm W$ scaling relation. Adopting a larger but still plausible value, such as $f=10$, would increase the inferred jet power and shift the spin estimate toward the moderately/rapidly spinning regime.

The accretion rates may also be overestimated due to the limited spatial resolution of SDSS imaging and the scatter in empirical BH mass--host luminosity relations. The alternative estimates in Section~\ref{sect:rate2} generally give lower accretion rates, which would in turn imply higher spins. Therefore, while our fiducial calculation favors moderate spins, values $a\gtrsim0.5$ remain plausible once the dominant systematic uncertainties are properly accounted for.

\begin{figure}
    \centering
    \includegraphics[width=0.6\textwidth]{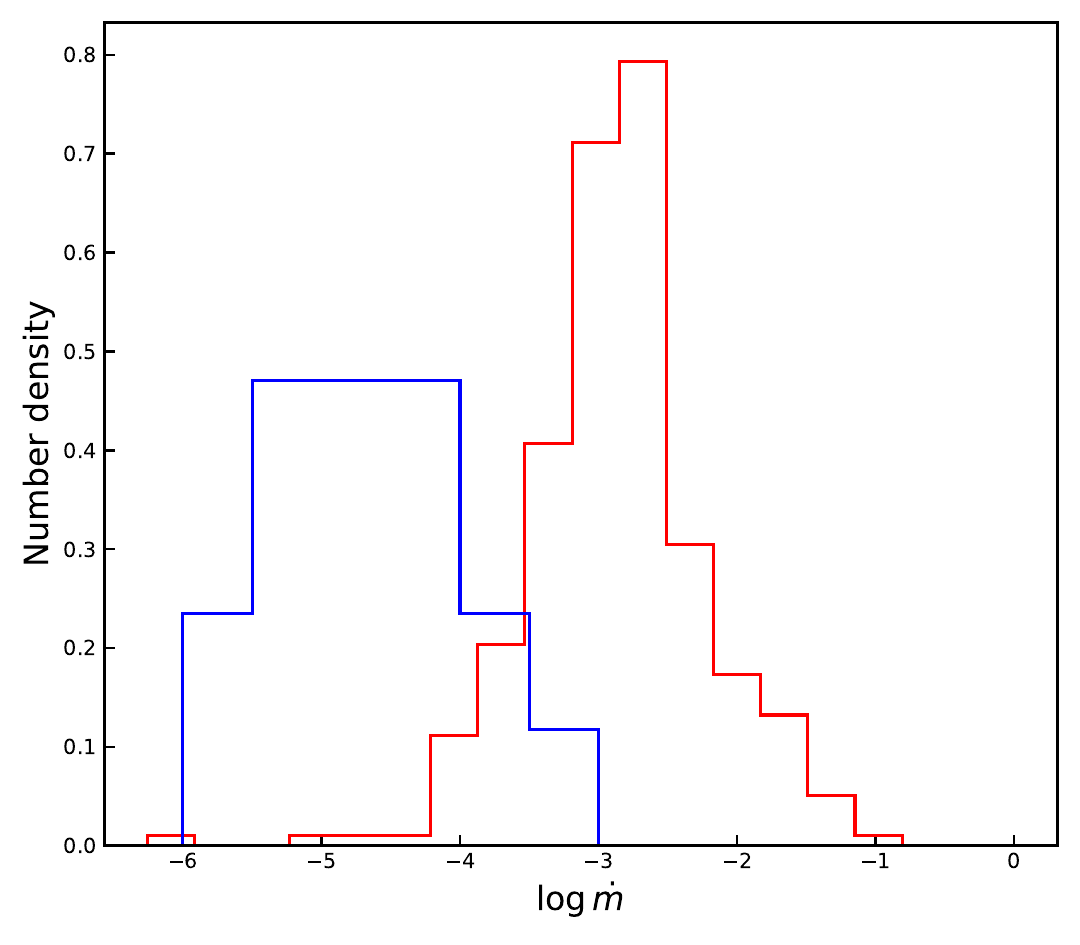}
    \caption{
    Comparison of Eddington ratio distributions between our LoTSS sample (red) and the 3CR FR I sample from \citet{he2024magnetically} (blue). The LoTSS sources exhibit systematically higher Eddington ratios, which may explain the slightly lower inferred black hole spins in our sample than in the brighter 3CR population.}
    \label{fig:compare_mdot}
\end{figure}

\section{Summary}

We present a comprehensive study of the accretion properties and jet powers of 289 FR I radio galaxies selected from the LoTSS Data Release 1. The main findings are as follows:

\begin{enumerate}
     \item The Eddington-scaled accretion rates span $-6.84 < \log \dot{m} < -0.87$ (median $\approx$ -2.84). The vast majority of sources lie below $\dot{m} = 0.01$, indicating that their central engines are well described by ADAFs. Independent estimates based on H$\beta$ and stellar velocity dispersion confirm the low-accretion interpretation.

    \item The radio-estimated jet powers (with $f=1$) are generally larger than the maximum values predicted by the Blandford–Znajek mechanism for standard ADAFs, even for a rapidly spinning black hole with $a = 0.95$. This demonstrates that the ADAF model significantly underestimated the jet power in the majority of our sample, implying stronger magnetic fields in the inner accretion flow.

    \item The MAD scenario, in which large-scale poloidal magnetic flux accumulates near the event horizon, fully accounts for the observed powerful jets. Within the MAD framework, the data favor moderately spinning black holes with $a<0.5$ under our fiducial assumptions. Larger values of the jet-power normalization factor $f$ or lower accretion-rate estimates would shift the inferred spins toward the moderately to rapidly spinning regime.

    \item Our LoTSS sample occupies a similar region in the $Q_{\rm jet}$--$M_{\rm BH}$ plane as the bright 3CR FR I sample studied by \cite{he2024magnetically}, indicating that MADs are common in FR I radio galaxies across a wide range of luminosities and that the MAD requirement is a general feature of this population.
\end{enumerate}

\begin{acknowledgements}

B.Y.\ is supported by NSFC grants 12322307, 12273026, and 12361131579; supported by “the Fundamental Research Funds for the Central Universities”; Xiaomi Foundation / Xiaomi Young Talents Program; The data analysis in this paper have been done on the supercomputing system in the Supercomputing Center of Wuhan University. 
MFG is supported by the National Science and Technology Major Project of China (No. 2024ZD1100601), the National Science Foundation of China (grant 12473019), the Shanghai Pilot Program for Basic Research-Chinese Academy of Science, Shanghai Branch (JCYJ-SHFY-2021-013), the National SKA Program of China (Grant No. 2022SKA0120102), and the China Manned Space Project with No. CMS-CSST-2025-A07.L.C. is supported National SKA Program of China
(2022SKA0120102), National Key R\&D program of China
(2024YFA1611403), and Shanghai Pilot Programme for Basic
Research, CAS Shanghai Branch (JCYJ-SHFY-2021-013).
\end{acknowledgements}

\section*{Data Availability}
The data used in this work are publicly available. The optical imaging and photometric data, as well as the spectroscopic data, are obtained from the Sloan Digital Sky Survey (SDSS,\url{https://www.sdss.org/} 
). The derived galaxy properties are taken from the MPA-JHU value-added catalogs (\url{http://www.mpa-garching.mpg.de/SDSS/}
).The radio observation is available in \cite{mingo2019}.

\bibliographystyle{raa}
\bibliography{example}

\label{lastpage}

\end{document}